# Energetic Versus Sthenic Optimality Criteria for Gymnastic Movement Synthesis

F. LEBOEUF, G. BESSONNET, P. SEGUIN and P. LACOUTURE

*Laboratoire de Mécanique des Solides, CNRS-UMR 6610, Université de Poitiers,*

*SP2MI, Bd. M. & P. Curie, BP 30179, 86960 Futuroscope Chasseneuil Cedex, France*

leboeuf@lms.univ-poitiers.fr ; bessonnet@lms.univ-poitiers.fr

**Abstract**

Dynamic synthesis of human movements raises the question of the selection of a suitable performance criterion able to generate proper dynamic behaviors. Two quite different criteria are likely to be appropriate candidates: the minimum effort cost (or sthenic criterion) and the minimum energy cost. The paper is aimed at clarifying the dynamic effects of these two fundamental criteria when considering movements executed with liveliness as they are in gymnastic. It is well known that the former cost generates movements with smooth dynamics. A special attention is devoted to the latter. The optimal control theory shows that minimizing the energy consumption results in actuating inputs of bang-off-bang type producing momentum impulses. When achieving dynamic synthesis, this criterion makes necessary to account for bounds set on driving torques. Moreover, when dealing with one-sided contacts, as in floor handstands, the unilaterality of contact forces must be explicitly accounted for since it tends to be infringed by impulsive efforts.

Numerical simulations of these formal properties are carried out using a parametric optimization technique, and considering the raising phase of floor handstands. It is shown that the energetic criterion tends to generate movements which exhibit similarities with their real counterpart performed by an expert gymnast. Conversely, the sthenic criterion produces movements quite different. But, a salient fact is that these ones proved to be easier to perform by young beginners. As a result, they could help to coach novice gymnasts.

## 1. Introduction

Dynamic analysis of human movement has been extensively developed during the last three decades. The central problem is the determination of internal forces, especially the driving joint torques which control and give momentum to the motion. Two main approaches are at stake: inverse dynamics and optimal dynamic synthesis.

Due to the fact that internal efforts can not be directly assessed, inverse dynamics is an essential approach to be used to deal with human movement analysis. It needs complete information on motion kinematics which is the key point to be dealt with in this case. A video recording system provides, according to an acquisition frequency, a sequence of positions of anatomic markers. Using data filtering together with numerical differentiation techniques allows velocities and accelerations to be derived from recorded positions. It should be noted that both measurement and data processing hold uncertainties. The first suffers mainly from its lack of accuracy for identifying relative positions of joint rotation axes and centers of rotation. The second is very likely to produce discrepancies between real velocities and accelerations and their computed counterparts of which maximum values generally truncate the real ones. Consequently, the evaluated kinematic parameters could provide, through an inverse dynamics model, joint actuating torques with values fairly different from the real ones. Nevertheless, the inverse dynamics approach generally gives valuable results representing a basic assessment of internal efforts which were at work to generate the recorded movement. It can help to understand the joint motion coordination as during normal and pathological gait [1-3]. It is helpful also to analyze ergonomic conditions, especially while performing load lifting [4-5]. Another objective may consist in identifying some performance parameters when considering athletic exercises [6-8].

Unlike inverse dynamics, optimal dynamic synthesis needs either few experimental data, or not at all. However, it requires a great amount of computational effort. The movement to be dealt with is generated using an optimization technique which minimizes some performance criterion. The method consists in extracting an optimal solution from the equations governing the motion dynamics, while satisfying typical constraints the movement must comply with. This approach is not dependent on experimental uncertainties, and it correlates at best kinematics with forces that create the movement. Consequently, it could be an alternative to the inverse dynamics method. However, the minimizing criterion could generate, according to its nature, motions noticeably different from related human movements. The problem is that we do not know really what dynamic criterion governs human motricity. Thus, the choice of a performance criterion is simply a working basis. Its relevance will help to generate movements with kinematics and dynamics having insightful similarities with their real counterparts.

In this way, a variety of motion-synthesis attempts were carried out in order to get better knowledge of human movement dynamics. Two general methods have been used: optimal control theory and parametric optimization. The first method was implemented in [9-11] using the Pontryagin Maximum Principle (PMP) to study human locomotion [9] and for generating human-like gait of a planar biped [10, 11]. The PMP is a powerful mathematical tool which gives formally exact optimal solutions. However, its implementation may present some intricacy when dealing with one-sided contacts and considering multibody systems having a great number of degrees of freedom. This double situation is encountered when taking into consideration complex movements such as complete gait cycles and various gymnastic exercises. Parametric optimization makes it possible to cope with such dynamic and kinematic aspects of movements to be generated. This approach was developed on the basis of quite different intersegmental actuating models and related cost functions.

Authors like Pandy et al. [12,13], Anderson and Pandy [14], and Eberhard et al. [15] considered musculoskeletal models accounting for distinctive mechanical and physiological characteristics of each muscle, to generate human jumping [12,15], sit-to-stand movements [13] and walking cycles [14]. Various performance criteria were taken into account according to the movement considered: minimal time integral of square muscle forces [12,13] and their time derivatives [13], minimal task-time [12], minimal metabolic energy expended per unit distance covered [14]. Movement synthesis based on musculoskeletal models is aimed at revealing neuromuscular coordination. Yet, this approach suffers from high complexity modeling together with uncertainties about the mechanical and physiological muscle behavior. Although this method deals with essential aspects of human movement organization, it is not very well adapted for generating wide-range movements with strong dynamic effects.

If one wants to optimize efficiently the global dynamics of complex movements, a simple joint intersegmental actuating model is much easier to handle than its musculoskeletal counterpart. This approach was particularly used for generating gait cycles designed for bipedal walking [16-20] as well as for quadrupedal locomotion [21]. Kuželički et al. [22] achieved the synthesis of sit-to-stand movements for trans-femoral amputees with leg-prosthesis and for normal persons using a dynamic model comprising 11 degrees of freedom. The cost functional to be minimized consisted of two main terms representing the integrals of both joint actuating torques and their time derivatives. As shown in [10], the latter term has the effect of smoothing the driving torque variations, which is a necessary precaution for persons fitted with prosthesis. The performance criterion generally used for generating gait cycles is the minimum-effort cost. Since during gait the body segments are essentially submitted to gravity, the minimization of joint driving torques tends to generate upright walking patterns which require moderate actuating efforts to counterbalance the gravity effects [19, 20]. However, Chevallereau and Aoustin [18], and Muraro et al. [21] made also use of an energy cost to generate walking and running cycles for simplified bipeds. In fact, in section 3, we will recall that such a criterion generates bang-off-bang actuating inputs, which raises some questions regarding its application to human movement synthesis performed through a parametric optimization technique.

Indeed, the present paper is specifically aimed at showing that the minimum-effort cost and the minimum-energy cost yield movements exhibiting quite different kinematic features and dynamic characteristics. Using an accurate state parameterization, we will also emphasize that dealing with an optimal control problem using a parametric optimization technique produces suboptimal solutions, owing to the fact that unknown functions defined along the motion time are characterized by a finite number of optimized discrete parameters. Moreover, approximating state functions using high order polynomials as in [16-18, 21], or splines of class $C^n$, $n \geq 2$ [19, 20], imply the continuity of joint accelerations. As a result, optimized joint driving torques will be continuous as well, whatever the minimized criterion may be. Especially, instead of bang-off-bang optimal controls, the energy criterion will produce continuous actuating inputs, even though they are not subjected to numerical bounds, which has no mathematical sense from the optimal control theory viewpoint [23]. This remark makes necessary the introduction of explicit limits for the driving torques in order to find relevant suboptimal solutions.

An outline of the next sections is as follows. In the section 2, the content of the optimal-dynamics problem we intend to deal with is detailed. We explicitly show in section 3 how the energetic cost we want to implement generates bang-off-bang actuating torques. Next, in section 4, the original dynamic optimization problem is converted into a minimization problem of mathematical programming using a state-parameterization method. Numerical simulations are presented in section 5. They were carried out considering a basic gymnastic movement, the handstand, performed as a floor exercise.

## 2. Stating a dynamic optimization problem

Optimal motion synthesis is based on stating and solving a dynamic optimization problem. This can be dealt with as an optimal control problem, or converted into a parametric optimization problem. We will briefly consider the former problem in Section 3 in order to gain an insight into the formal nature of optimal control inputs which result from the criterion minimized, and the latter will be developed in Section 4 to compute suboptimal solutions using standard computing codes implementing quadratic sequential programming algorithms.

As regards its kinematics and dynamics, the human body is commonly described as a multibody system made of rigid segments and actuated rotational joints. In Section 5, a gymnast is depicted as a rooted planar system with an open tree-like topology (Fig. 3).

*2.1. Dynamics equations and boundary conditions*

The dynamics of movements we consider over an interval of time $[0,T]$ is governed by a set of equations summarized as the single *n*-vector relationship

$$t \in [0,T], \; B(q(t), \dot{q}(t), \ddot{q}(t)) = \tau(t), \tag{1}$$

where

- $q = (q_1, ..., q_n)^T$ is the vector of joint coordinates $q_i$, $i \leq n$, $n$ being the number of degrees of freedom.
- $\dot{q}$ and $\ddot{q}$ are the first and second order time-derivatives of $q$.
- $\tau = (\tau_1, ..., \tau_n)^T$ is the vector of joint driving torques, assuming that all degrees of freedom are actuated.

The phase trajectory $t \to (q(t), \dot{q}(t))$ must comply with some initial and final conditions we formally write under the form

$$\begin{cases} \varphi_0(q(0), \dot{q}(0)) = 0 \, (\in \Re^{n_0}, n_0 \leq 2n) \\ \varphi_T(q(T), \dot{q}(T)) = 0 \, (\in \Re^{n_T}, n_T \leq 2n) \end{cases}. \qquad (2)$$

If $n_0 < 2n$, or $n_T < 2n$, then, the initial state $(q(0), \dot{q}(0))$, or the final state $(q(T), \dot{q}(T))$, is incompletely specified or subjected to constraints described by the function $\varphi_0$ or $\varphi_T$. In the example dealt with in section 5, the initial state is constrained ($n_0 < 2n$) while the final state is specified ($n_T = 2n$).

*2.2. State constraints*

Generally, joint motions have limited range and must avoid hyperextension, which results in formulating inequality constraints such as

$$t \in [0, T], k \leq n, \begin{cases} g_k(q(t)) := q_k^{\min} - q_k(t) \leq 0 \\ g_{n+k}(q(t)) := q_k(t) - q_k^{\max} \leq 0 \end{cases}. \qquad (3)$$

Let us notice that the joint velocities could be submitted to similar limitations.
We formally combine inequalities (3) into the vector function

$$g = (g_1, ..., g_{n_g})^T, \, t \in [0, T], \, g(q(t)) \leq 0, \qquad (4)$$

where $n_g$ stands for the total number of constraints. It should be noted that $q_k^{\min}$ and $q_k^{\max}$ represent given data (see Table 1 in section 5.2).

*2.3. Sthenic constraints*

In addition, joint driving torques must be bounded too, especially when minimizing an energetic cost (see section 3). Consequently, there is the need for taking into account limiting constraints as

$$t \in [0, T], \, k \leq n, \begin{cases} h_k(\tau(t)) := \tau_k^{\min} - \tau_k(t) \leq 0 \\ h_{n+k}(\tau(t)) := \tau_k(t) - \tau_k^{\max} \leq 0 \end{cases}, \qquad (5)$$

where the lower bounds $\tau_k^{\min}$ are assumed to be negative. Generally, such limits are not symmetric: $\tau_k^{\min} \neq -\tau_k^{\max}$.

Moreover, interaction forces such as ground reactions due to one-sided contacts must fulfill unilaterality requirement and non sliding conditions (see section 5). Constraints of this type depend explicitly on phase variables and actuating torques. Thus, they can be written as

$$t \in [0,T], j \leq n_r, h_{2n+j}(q(t), \dot{q}(t), \tau(t)) \leq 0, \tag{6}$$

$n_r$ being the number of constraints.

We will term *sthenic constraints* the conditions (5) and (6) that we summarize using the vector function

$$h = (h_1,...,h_{n_h})^T; t \in [0,T], h(q(t), \dot{q}(t), \tau(t)) \leq 0. \tag{7}$$

The subscript $n_h$ represents the number of scalar constraints.

*2.4. Performance criteria*

We consider the following two criteria
- A *sthenic cost* defined as the time integral of quadratic driving torques

$$J_s = \frac{1}{2} \int_0^T \tau(t)^T D_s \tau(t) dt, \quad D_s = \text{diag}(\zeta_1,...,\zeta_n), \tag{8}$$

- An *energetic cost* which is the time integral of absolute joint actuating powers

$$J_e = \int_0^T \sum_{i=1}^n \xi_i |\dot{q}_i(t)\tau_i(t)| dt, \tag{9}$$

where $\zeta_i$ and $\xi_i$ are weighting coefficients.

In fact, we will take into consideration a combination of these two costs, that is to say the mixed criterion

$$J = \alpha J_s + (1-\alpha) J_e, \quad \alpha \in [0,1] \tag{10}$$

which gives $J_s$ and $J_e$ when $\alpha$ takes the values 0 and 1 respectively.

Finally, the dynamic optimization problem can be summarized as follows: a value of α being given, find a joint trajectory $t \to q(t)$ and a vector-function $t \to \tau(t)$ solution of (1), minimizing the criterion (10) while satisfying the boundary conditions (2), together with the double set of constraints (4) and (7). This problem is dealt with in the following two sections.

## 3. Optimal actuating inputs

The analytic nature of optimal driving torques minimizing the criterion (10) can be revealed using the Pontryagin Maximum Principle [23] without having to solve the optimal control problem. We refer the reader to [10, 24] for a general approach developed to deal with dynamic optimization problems related to multibody systems. It is shown that the necessary conditions for optimality are easier to derive from a Hamiltonian dynamic model than from its Lagrangian counterpart.

Thus, let us briefly introduce:

- The Lagrangian: $L(q,\dot{q}) = T(q,\dot{q}) - V(q)$, $T$ being the kinetic energy and $V$ the gravity potential of the mechanical system.
- The conjugate momentum: $p = \partial L/\partial \dot{q}$.
- The Hamiltonian vector phase variable: $x = (q^T, p^T)^T$.

Then, it can be easily shown [10] that the Hamilton equations equivalent to (1) take the simple form of the $2n$-order vector-state equation

$$\dot{x} = F(x) + A\tau, \qquad (11)$$

where $A = \begin{pmatrix} 0_{n \times n} \\ I_{n \times n} \end{pmatrix}$.

Moreover, through (8) and (9), the criterion $J$ can be written under the standard form

$$J = \int_0^T l(x(t), \tau(t)) dt. \qquad (12)$$

Then, defining the Pontryagin function [23]

$$y \in \Re^{2n}, \ H(x, y, \tau) = y^T (F(x) + A\tau) - l(x, \tau), \qquad (13)$$

and putting aside momentarily the state constraints (4), the Pontryagin Maximum Principle states that for any solution $(x, \tau)$ of (11) satisfying the constraints (2) and (7), and minimizing (12), there exists a $2n$-vector function $t \to y(t)$ satisfying the so-called adjoint system

$$t \in [0, T], \ \dot{y} = -(\partial H/\partial x)^T, \qquad (14)$$

and the maximality condition for $H$

$$t \in [0, T], \ H(x(t), y(t), \tau(t)) = \underset{u \in U}{Max} \ H(x(t), y(t), u), \qquad (15)$$

where $U$ stands for the set of feasible control variables $\tau_i s$ defined by the sthenic constraints (5) and (6). If only the constraints (5) are taken into account, the set $U$ appears as a parallelepiped in the $n$-dimensional control space. Constraints (6) eliminate half-spaces (see [24]) which truncate $U$ and reduce it to a polyhedron.

We need to examine the maximality condition (15) when the only constraints (5) are accounted for. Knowing the expression of the Lagrangian $l$ in (12) and (13) through (8) and (9), the condition (15) allows the optimal inputs $\tau_i s$ to be explicitly formulated as functions of $x$ and $y$. When $J$ is reduced to $J_s$, it is well known that optimal inputs appear as saturating functions which are continuous functions truncated by upper and lower bounds of the $\tau_i s$ (see [23]). At this point, we want to be more specific about the optimal inputs which result from the purely energetic cost we intend to implement as the limiting case $\alpha = 0$ in (10).

First, from (11), we have $\dot{q}_i \equiv \dot{x}_i = F_i(x)$, $i \leq n$. Then, the Lagrangian of $J_e$ in (9) can be written as

$$l_e(x,\tau) = \sum_{i=1}^{n} \xi_i |\tau_i F_i(x)|.$$

Accordingly, the Pontryagin function takes the form

$$H(x,y,u) = C(x,y) + \sum_{i=1}^{n} (y_{n+i} u_i - \xi_i |F_i||u_i|),$$

in which $C(x,y)$ stands for terms independent of $u$.

This relationship shows that maximizing $H$ amounts to maximizing the sum in the right hand member. This maximization problem can be solved explicitly when the only active sthenic constraints are reduced to (5), i.e. when the set $U$ is a parallelepiped. In that case, it is possible to write

$$\underset{u \in U}{Max} \sum_{i=1}^{n} (y_{n+i} u_i - \xi_i |F_i||u_i|) = \sum_{i=1}^{n} \underset{\tau_i^{min} \le u_i \le \tau_i^{max}}{Max} (y_{n+i} u_i - \xi_i |F_i||u_i|), \quad (16)$$

because the $u_i$s vary in $U$ independently of each other.

Next, the problem to be solved consists of finding a real variable $u_i$ solution of the elementary maximization problem

$$\underset{\substack{\tau_i^{min} \le u_i \le \tau_i^{max} \\ \alpha_i \in \Re, \beta_i \ge 0}}{Max} (\alpha_i u_i - \beta_i |u_i|). \quad (17)$$

As shown in the appendix, the solution can be expressed using a *dead zone function* and the *sign function*. Since $u_i$ is determined for every fixed time $t$, we have $\tau_i(t) = u_i$. Consequently, from the relationships (A2) and (A3), we get

$$\tau_i(t) = \begin{cases} \text{dez}\left(\dfrac{y_{n+i}(t)}{\xi_i |F_i(x(t))|}, \tau_i^{min}, \tau_i^{max}\right), & \text{if } F_i(x(t)) (\equiv \dot{q}_i(t)) \ne 0 \\ \tau_i^{max} \text{sign}(y_{n+i}(t)), & \text{if } F_i(x(t)) = 0 \end{cases}. \quad (18)$$

The figure 1 shows typical variations of an optimal actuating torque resulting from the energy cost minimization. It must be emphasized that whatever the limiting values $\tau_i^{min}$ and $\tau_i^{max}$ may be, the corresponding optimal input will be of bang-off-bang type as shown in Figure 1. If limiting values are not explicitly taken into account, singular solutions will result, taking the zero value over the entire time interval except for discrete points where the $\tau_i$s will generate infinite impulses. However, optimization carried out using state-parameterization techniques has the effect of smoothing the suboptimal solutions which result. Thus, suboptimal actuating inputs are smoothed as well. Nevertheless, if they are not subjected to predefined bounds, they will be likely to reach disproportionate values over brief time-subintervals. This situation must be avoided.

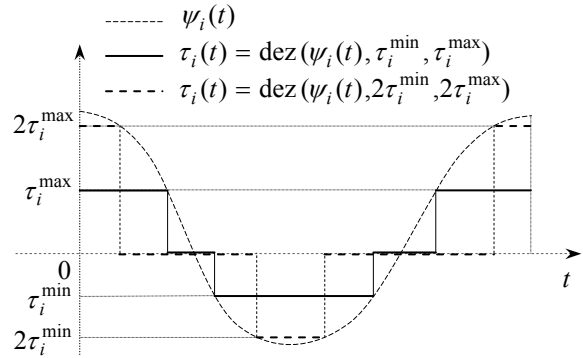

Fig. 1. *Two examples of bang-off-bang optimal input $\tau_i$ minimizing the energy cost, according to the values of lower and upper bounds to be respected.*

To finish, let us add that when the constraints (6) are accounted for, optimal inputs are no longer of purely bang-off-bang type since the set $U$ is a truncated parallelepiped. However, the problem is not fundamentally changed, because as soon as constraints in (6) will not be active, the bang-off-bang nature of optimal inputs will reappear.

## 4. Parametric optimization method

Parameterizing an optimization problem consists first in approximating some variables to be optimized using a finite set of discrete parameters and, second, in recasting the criterion to be minimized into a simple function of these parameters. The reader is referred to [25] for a short presentation of general ideas commonly used to convert optimal control problems into parameter optimization problems. Practical methods for solving such problems could be found in [26]. Considering a dynamic optimization problem, the parameterization may be carried out according to quite different approaches depending on whether state variables or control variables are parameterized using approximation functions.

The latter method was used notably in [12-14] where the values of control variables defined at a finite number of points along the motion time are taken into account as optimization variables. The control history is then reconstructed by linear interpolation between nodal points. Then, an initial state being given, a forward integration of motion equations yields a generalized final state together with the value of the cost function with respect to the optimization variables. The parameterization of the original problem is thus completed. This technique is appealing because it eliminates untimely oscillations of state variables between knots. However, its accuracy strongly depends on a given initial state together with the precision of the forward integration technique which could be very time consuming. It does not seem suitable for achieving the optimal synthesis of movements of which end states need to be optimized as required, for example, for a gait cycle [19, 20].

Parameterizing the state variables needs to use smooth approximation functions, at least twice differentiable in order to ensure the existence of accelerations. This preliminary operation being done, actuating inputs can be easily computed by performing inverse dynamics. The conversion of the original dynamic optimization problem into parametric optimization is straightforwardly achieved by the computation of the cost function with respect to the new optimization variables.

A number of approaches have been used to approximate the state variables which can be chosen in the task space to describe, for instance, swing foot and hip trajectories for generating gait steps [16, 17] or, more generally, in the joint space in order to be in command of the joint motion coordination. In either cases, the most frequently used approximation functions are polynomials [16-18, 21] and splines [19, 20]. The former are easy to use but are prone to undesirable oscillations, especially if they have high order. This could result both in a limited minimizing effect of the cost function, and in sizeable constraint infringements. Spline functions represent a more accurate choice, either under the form of B-splines used as base functions [27, 28], or low order polynomials successively linked to each other at connecting points (or knots) defined along the motion time [19, 20]. In the latter case, polynomial functions of order three are sufficient to ensure the existence of second order derivatives (i.e. accelerations) at connecting points. This way, one gets spline functions of class $C^2$ over the traveling time. However, the non differentiability of accelerations at knots results in jerky variations which can be transferred to actuating torques. A more satisfactory approach developed in [20] consists of using splines of class $C^3$ obtained by connecting four order polynomials up to their third derivatives. Rough variations of actuating torques at knots are thus avoided. We have implemented this more accurate technique in order to generate smooth movements on the overall motion time.

4.1. *Defining a set of approximation functions*

We adapt and summarize the method developed in [29]. First, a set of connecting times must be defined over the interval [0, $T$] as

$$\{t_1(=0),...,t_j,...,t_{N+1}(=T)\}, \ t_{j+1} - t_j = T/N,$$

where $N$ is the number of subintervals of equal length.

Second, over each subinterval $I_j := [t_j, t_{j+1}]$, and for every joint coordinate $q_i$, a four order polynomial $P_{ij}$ is defined over the normalized interval [0,1] such that

$$j \leq N, \begin{cases} t \in I_j, \tau = \dfrac{t - t_j}{t_{j+1} - t_j}, \\ q_i(t) \approx P_{ij}(C_{ij}, \tau) = c_{ij0} + c_{ij1}\tau + c_{ij2}\tau^2 + c_{ij3}\tau^3 + c_{ij4}\tau^4 \end{cases}, \quad (19)$$

where $C_{ij}$ is the 5-order vector of coefficients: $C_{ij} = (c_{ij0}, c_{ij1}, c_{ij2}, c_{ij3}, c_{ij4})^T$.

Next, for every subscript *i*, the polynomials $P_{ij}$ are successively linked to each other, up to their third derivatives, as follows

$$2 \leq j \leq N, \begin{cases} P_{i,j-1}(1) = P_{ij}(0) \\ \dot{P}_{i,j-1}(1) = \dot{P}_{ij}(0) \\ \ddot{P}_{i,j-1}(1) = \ddot{P}_{ij}(0) \\ \dddot{P}_{i,j-1}(1) = \dddot{P}_{ij}(0) \end{cases} \quad (20)$$

Furthermore, we set

$$\begin{cases} P_{i1}(0) = q_i(0) \\ \dot{P}_{i1}(0) = \dot{q}_i(0) \\ P_{iN}(1) = q_i(T) \\ \dot{P}_{iN}(1) = \dot{q}_i(T) \end{cases}, \quad 2 \leq j \leq N, \quad P_{ij}(0) = q_i(t_j). \quad (21)$$

Relationships in (21) mean that the end states $(q_i(0), \dot{q}_i(0))$ and $(q_i(T), \dot{q}_i(T))$, together with inner values of the $q_i s$ at connecting points will be considered as free parameters to be optimized, which we formally assemble in the set of vectors

$$i \leq n, \; X^i = (x_1^i, ..., x_{N+3}^i) := (\dot{q}_i(0), q_i(0), ..., q_i(t_j), ..., q_i(T), \dot{q}_i(T))^T. \quad (22)$$

Now, using the relationships (20) and (21), the coefficients in (19) can be computed as linear functions of the above parameters. Actually, for any subscript *i*, there are 5*N* coefficients introduced in (19). They are subjected, through (20) and (21), to 5*N*-1 linear equations. Reducing to three the order of one of the polynomials $P_{ij}$, for instance the last, $P_{iN}$, by setting $c_{iN4} = 0$, the set of equations (20) and (21) can be solved in the 5*N*-1 remaining coefficients that we bring together in the vector $C^i = (C_{i1}^T, ..., C_{iN}^T)^T$. Thus, this one appears as a function of the vector $X^i$ defined in (22), that is

$$C^i = \Phi(X^i). \quad (23)$$

It should be noted that the function $\Phi$ is formally the same for every index *i*. Accounting for (23) in (19), each polynomial $P_{ij}$ can be recast as a function $\varphi_{ij}$ such as

$$P_{ij}(C_{ij}, \tau) = \varphi_{ij}(X^i, \tau).$$

Next, defining a function $\phi_i$ over the time interval [0, *T*] by setting

$$\forall j \leq N, \; t \in I_j, \; \phi_i(X^i, t) = \varphi_{ij}(X^i, \tau),$$

the generalized coordinate $q_i$ is approximated by $\phi_i$:

$$t \in [0, T], \; q_i(t) \approx \phi_i(X^i, t). \quad (24)$$

Moreover, if we set

$$X = ((X^1)^T, ..., (X^n)^T)^T,$$

$$\phi(X,t) = (\phi_1(X^1,t), ..., \phi_n(X^n,t))^T,$$

the configuration vector $q$ is approximated by the function $\phi$ as

$$t \in [0,T],\ q(t) \approx \phi(X,t). \tag{25}$$

4.2. *Stating a parametric optimization problem*

Through (25), the vector of actuating torques $\tau$ in (1) can be written as a vector $\tau^*$ depending on $X$ and $t$ as follows

$$\tau^*(X,t) := B(\phi(X,t), \frac{\partial \phi}{\partial t}(X,t), \frac{\partial^2 \phi}{\partial t^2}(X,t)).$$

$$\tau(t) \approx \tau^*(X,t)$$

The criteria $J_s$ and $J_e$ in (8) and (9) are then approximated by the functions of $X$

$$J_s \approx F_s(X) = \tfrac{1}{2} \int_0^T \tau^*(X,t)^T D_s \tau^*(X,t) dt, \tag{26}$$

$$J_e \approx F_e(X) = \int_0^T \tau^*(X,t)^T D_e \frac{\partial \phi(X,t)}{\partial t} dt. \tag{27}$$

The mixed criterion (10) becomes

$$J \approx F(X) = \alpha F_s(X) + (1-\alpha) F_e(X). \tag{28}$$

The boundary constraints (2) can be converted as

$$\varphi_0(\phi(X,0), \frac{\partial \phi(X,0)}{\partial t}) = 0$$

$$\varphi_T(\phi(X,T), \frac{\partial \phi(X,T)}{\partial t}) = 0 \tag{29}$$

where it should be seen that, through (25) and (22)

$$\frac{\partial \phi(X,0)}{\partial t} = (x_1^1,...,x_1^n)^T,\ \phi(X,0) = (x_2^1,...,x_2^n)^T,$$

$$\phi(X,T) = (x_{N+2}^1,...,x_{N+2}^n)^T,\ \frac{\partial \phi(X,T)}{\partial t} = (x_{N+3}^1,...,x_{N+3}^n)^T.$$

Now, the constraint functions $g$ and $h$ in (4) and (7) can be recast as the following functions of $X$ and $t$

$$g^*(X,t) = g(\phi(X,t), \phi_{,t}(X,t)),$$

$$h^*(X,t) = h(\phi(X,t), \phi_{,t}(X,t), \tau^*(X,t)).$$

Finally, the distributed constraints (4) and (7) will be accounted for as the discrete conditions set at connecting times

$$k \leq N + 1, \begin{cases} g^*(X, t_k) \leq 0 \\ h^*(X, t_k) \leq 0 \end{cases}. \qquad (30)$$

To conclude, the original dynamic optimization problem defined in (10), (2), (4) and (7) is converted into a standard minimization problem which consists of finding a discrete set *X* of optimization variables minimizing the cost function (28) ((26) or (27) as well) while satisfying equality constraints (29) and inequality constraints (30). This constrained non linear problem of mathematical programming can be solved using existing computing codes implementing sequential quadratic programming algorithms which have proved to be quite efficient.

**5.  Dynamic synthesis of a gymnastic movement: the handstand**

The handstand may be executed at all apparatus such as rings or high bar. As a gymnastic floor exercise (Fig. 2), it can be divided into three phases. First, the gymnast puts his hands in floor support. Second, after feet takeoff, he reaches the inverse balance. Third, the gymnast must perform a stationary handstand. This movement is considered as a fundamental gymnastic sequence which is difficult to perform by young beginners as it requires a specific development of muscular coordination. For this reason, it held researchers' attention. For example, in [30, 31], the authors focused their analyses on kinematic parameters and reaction forces measured using a video recording system and a force plate, respectively. In [32], the computation of joint actuating torques was carried out to complete the analysis of the motionless final posture. The authors used a simple dynamic model with three degrees of freedom, considering both legs and arms together. Through such analyses, one cannot know how the initial momentum of the raising phase and its segmental coordination influence the dynamic control of the handstand.

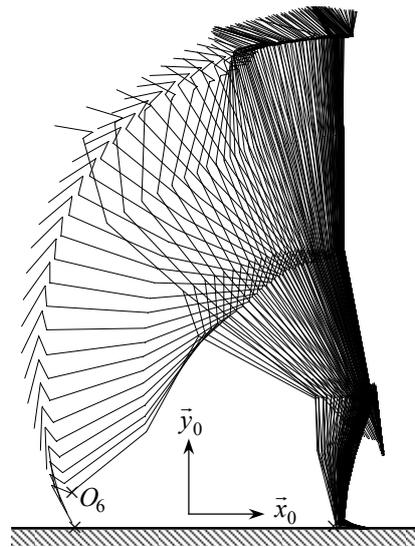

Fig. 2. *Stick diagram of a floor-handstand raising phase performed by an expert gymnast.*

In this section, our purpose is to achieve the dynamic synthesis of the raising phase of floor handstands in order to provide a new insight into the way the joint motions can be coordinated. An essential further objective of the paper is to show that minimizing a sthenic cost or an energetic cost as defined in section 3, generates movements with distinctive features that highlight the effects of each criterion and which have proved to be useful in coaching young gymnasts.

5.1. *Kinematic modeling and state constraints*

Through the raising phase of the handstand, the gymnast is modeled in his sagittal plane as a 9-link multibody system (Fig. 3). His hands are flat on the ground (link $L_0$), and both arms are assumed to perform the same movement (links $L_1$ and $L_2$). Trunk and head are considered as an only rigid link ($L_3$). As schematized in Fig. 3, the model has nine active joints.

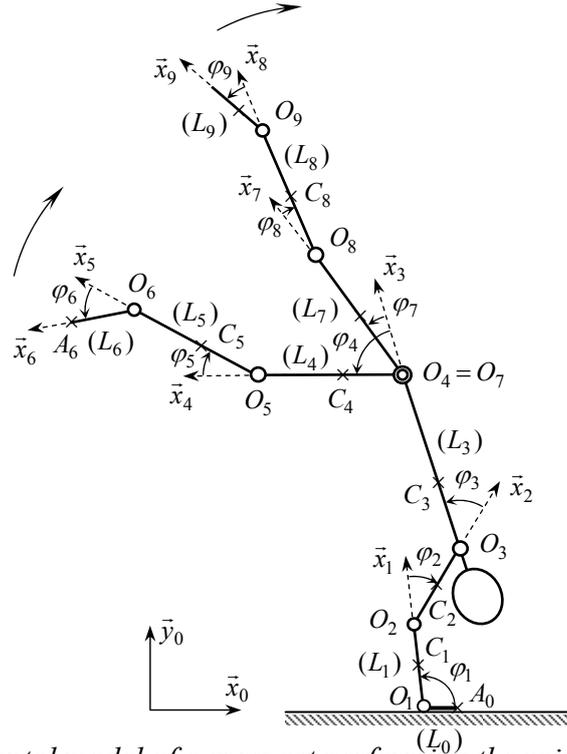

Fig. 3. *Planar segmental model of a gymnast performing the raising phase of a handstand on horizontal ground.*

The kinematic model is described by a set of nine generalized coordinates which are the joint angles of relative rotations between adjacent links. Thus, we introduce the configuration vector $q = (q_1,...,q_i,...,q_n)^T$ with $n = 9$ and $q_i = \varphi_i$ as defined in Fig. 3.

At initial time of the raising phase, the foot of the lower leg takes off. Only three equality constraints in (31) specify the initial state of the gymnast: the first indicates that the tip $A_6$ of the foot is still in contact with the supporting ground (Fig. 2), while the next two ones mean that its initial velocity is zero.

$$\begin{cases} \varphi_{0,1}(q(0),\dot{q}(T)) \equiv \overrightarrow{O_1 A_6}(q(0)) \cdot \vec{y}_0 = 0 \\ \varphi_{0,2}(q(0),\dot{q}(T)) \equiv \vec{V}(A_6(q(0))) \cdot \vec{x}_0 = 0 \\ \varphi_{0,3}(q(0),\dot{q}(T)) \equiv \vec{V}(A_6(q(0))) \cdot \vec{y}_0 = 0 \end{cases} \quad (31)$$

As joint positions and velocities are not specified, we emphasize that the initial posture and the initial momentum of the simulated gymnast will be optimized.

At final time, the handstand is achieved as prescribed by the complete set of constraints (34) where the $q_{iT}s$ stand for given values.

$$i \leq n, \begin{cases} \varphi_{T,i}(q(T),\dot{q}(T)) \equiv q_i(T) = q_{iT} \\ \varphi_{T,n+i}(q(T),\dot{q}(T)) \equiv \dot{q}_i(T) = 0 \end{cases}. \quad (32)$$

The state constraints (4) are reduced to the limitation of joint motion ranges at knees and ankles, that is

$$\begin{cases} (q_5^{\min}, q_5^{\max}) \equiv (q_8^{\min}, q_8^{\max}) = (-30, 0) \text{ (deg.)} \\ (q_6^{\min}, q_6^{\max}) \equiv (q_9^{\min}, q_9^{\max}) = (20, 60) \text{ (deg.)} \end{cases}. \quad (33)$$

5.2. *Dynamic modeling and sthenic constraints*

We used the Lagrange equations which give to (1) the formal structure

$$M(q)\ddot{q} + C(q,\dot{q}) + G(q) = \tau, \quad (34)$$

where $M$ is the $(n \times n)$- inertia-matrix, $C$ is the $n$-vector of Coriolis and centrifugal terms, and $G$ stands for the gravity terms.

Bounds set on actuating torques in (5) are given in Table 1. Indicated values are simply based on preliminary optimization tests carried out using a small number of nodal points, exactly five.

| Joints | Wrists | Elbows | Shoulders | Hips | Knees | Ankles |
|---|---|---|---|---|---|---|
| $i$ | 1 | 2 | 3 | 4, 7 | 5, 8 | 6, 9 |
| $\tau_i^{\max}$ (Nm) | Flexion 2×40 | Flexion 2×40 | Extension 2×40 | Flexion 70 | Extension 90 | Flexion 50 |
| $-\tau_i^{\min}$ (Nm) | Extension 2×25 | Extension 2×40 | Flexion 2×50 | Extension 110 | Flexion 50 | Extension 90 |

Table 1. *Limiting values of actuating torques.*

Contact forces exerted by the ground on the hands can be represented by a system of three forces, two of them being vertical forces applied to the hand extremities, the third representing the tangential friction component acting along the axis $(O_1; \vec{x}_0)$ (Fig. 4). Unilaterality of contact and non sliding condition can be expressed as constraints of type (6), that is:

$$\begin{aligned} h_{2n+1}(q,\dot{q},\tau) &\equiv -N_{O_1} < 0 \\ h_{2n+2}(q,\dot{q},\tau) &\equiv -N_{A_0} < 0 \\ h_{2n+3}(q,\dot{q},\tau) &\equiv T - f \times (N_{O_1} + N_{A_0}) < 0 \end{aligned}, \quad (35)$$

where $f$ is a dry friction coefficient. These interaction forces were computed with respect to $(q,\dot{q},\tau)$ using Newton-Euler equations formulated for the whole system.

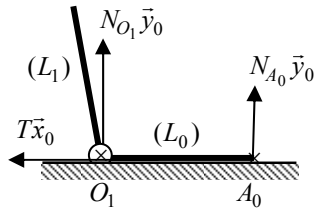

MUBO 05026      Final versionFig. 4. *System of forces exerted by the ground on the "hands"* ($L_0$).

## 5.3. *Numerical results*

We used the biometric data, given in table A1 of the appendix, of an expert gymnast.

A first group of simulations was carried out without accounting for the state constraints (33). Therefore, the legs are free to flex without limitations. Moreover, the relative position of the lower foot tip is not specified. Consequently, the initial hand-foot distance will be optimized. Results of nine simulations are shown in Table 2. They were obtained by varying two main parameters:

− The number $N$ of control points in (19) which is a fundamental data for the optimization technique employed. The actual values are 5, 10 and 20 which generate 72, 117 and 207 optimization parameters, respectively.

− The value of the weighting factor $\alpha$ in the mixed criterion (28) which is an essential characteristic of the minimization problem to be solved.

Four values of $\alpha$ were taken into account. The first, equal to 1, defines a purely sthenic criterion. The second value, and particularly the third, represents a nearly energy cost. The fourth which defines a purely energetic criterion was used only for the second and third values of $N$ which provide the better accuracy.

| $\alpha$ \ $N$ | Minimal cost $J$ (adimensional) | | | Actuating effort $J_s$ (adimensional) | | | Actuating work $J_e$ (J) | | | Initial foot-hand distance (m) | | |
|---|---|---|---|---|---|---|---|---|---|---|---|---|
| | 5 | 10 | 20 | 5 | 10 | 20 | 5 | 10 | 20 | 5 | 10 | 20 |
| 1.00 | 7.51 | 8.69 | 8.47 | 7.51 | 8.69 | 8.47 | 351 | 350 | 374 | 0.35 | 0.46 | 0.50 |
| 0.05 | 1.73 | 1.65 | 1.65 | 12.2 | 10.8 | 11.3 | 207 | 204 | 200 | 0.47 | 0.49 | 0.50 |
| 0.01 | 1.25 | 1.18 | 1.16 | 15.1 | 19.0 | 17.7 | 194 | 174 | 173 | 0.56 | 0.66 | 0.60 |
| 0.00 | | 0.97 | 0.95 | | 23.3 | 23.9 | | 169 | 165 | | 0.70 | 0.70 |

Table 2. *Minimal costs $J$, $J_s$, $J_e$, and optimal initial foot-hand distance with respect to the number $N$ of control points and the values of the weighting factor $\alpha$.*

On each of the first three lines in $\alpha$, the fluctuations of results versus the number $N$ of control points reflects the suboptimal nature of the minimization technique. On the columns, the variations are quite significant and express the change from the sthenic criterion to the energetic one. Results reveal that there are large increases of the total actuating effort, and at the same time the actuating work is greatly decreased. Fig. 5 shows three optimal movements computed for $N$ equal to 20. The corresponding time charts of actuating torques and contact forces are displayed in Fig. 6.

The change from the minimum effort criterion to the nearly minimum energy cost shows two different strategies for achieving the final balance of the handstand (Fig. 5): in the first case, both legs are initially subjected to important flexions and perform final extensions as soon as the center of mass of the gymnast reaches the vertical of the hands; in the third case, the upper leg does not flex at all while the lower leg makes a moderate flexion before its final extension, which limits the movement beyond the vertical. The first movement is very unlike the raising phase performed by the gymnast (Fig. 2). But, the third presents some similarities with the latter: the upper leg is kept in extension, and the lower leg movement beyond the vertical of the final posture is drastically reduced.

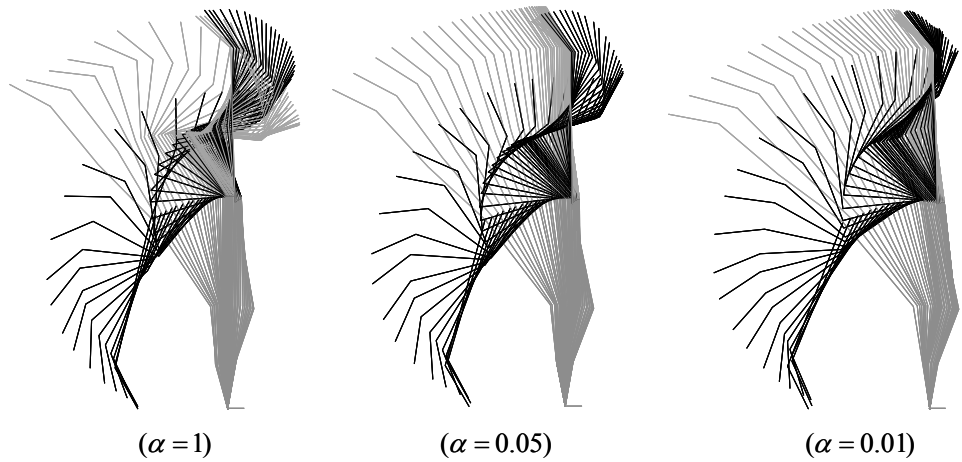

($\alpha = 1$)     ($\alpha = 0.05$)     ($\alpha = 0.01$)

Fig. 5. *Stick diagrams of optimal motions computed on the basis of three different criteria defined by the values of the weighting factor $\alpha$.*

In Fig. 6, actuating torques and contact forces show very different variations along the motion time according to the value of the weighting factor $\alpha$. The sthenic criterion ($\alpha = 1$) generates moderate driving torques exhibiting few variations in comparison with their quasi-energetic counterparts ($\alpha = 0.01$). In the latter case, joint torques do not cease oscillating between the zero value and their maximum authorized values. This denotes a tendency for bang-off-bang dynamic behavior, as shown in section 3. Accordingly, contact forces are also oscillating. It should be noted that the unilaterality requirement is satisfied over the whole time interval. However, the bound constraint set on the lower leg hip-torque is slightly infringed twice between control points at beginning of the movement.

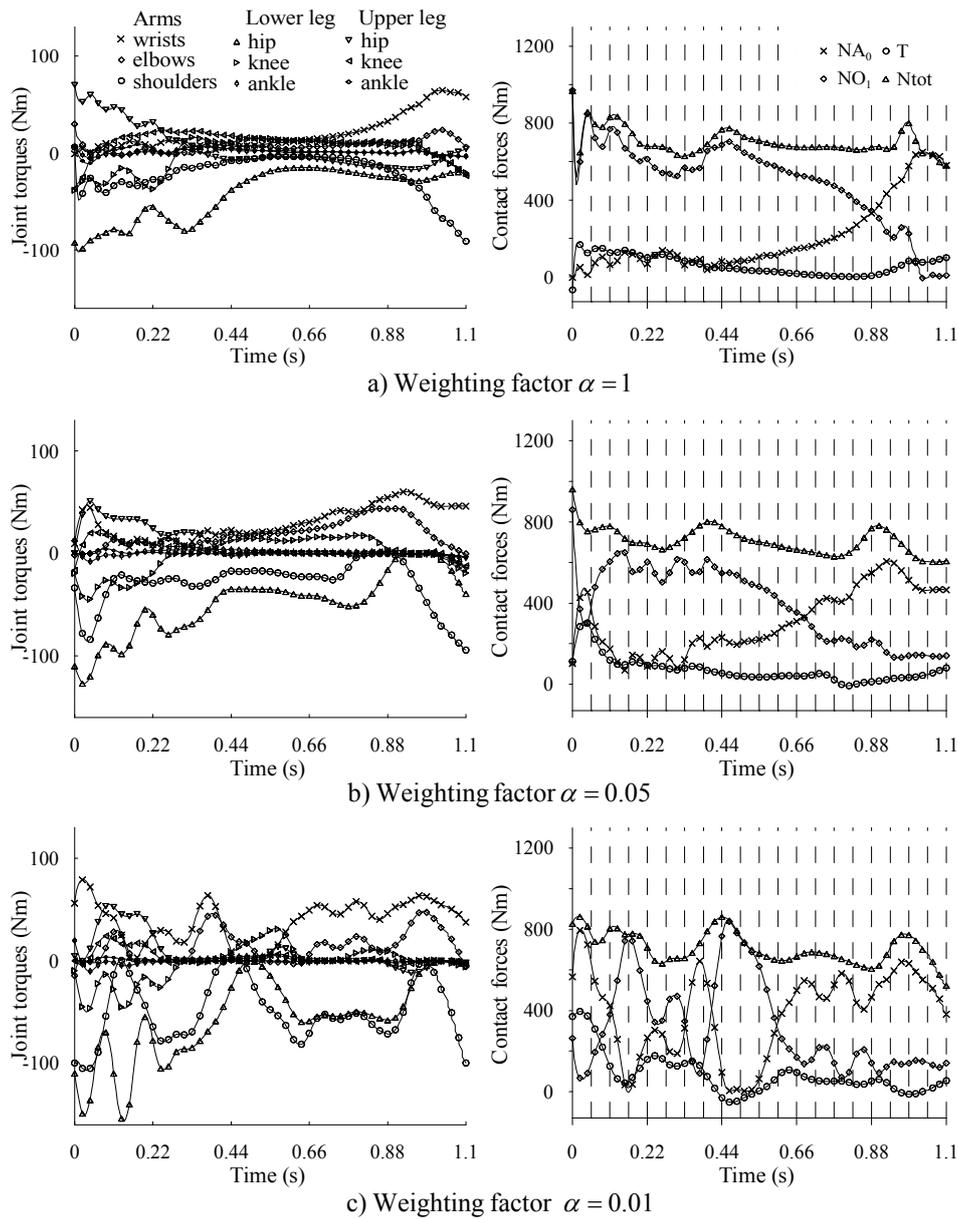

Fig. 6. *Time charts of actuating torques and contact forces for three values of the weighting factor $\alpha$; vertical hatched lines are drawn at control points.*

The purely energetic case ($\alpha = 0$) is shown apart in Fig. 7. The movement is wider than the previous ones. The initial hand-foot distance has increased while the actuating work has decreased (see Table 2). Unfortunately, the unilaterality condition is frequently infringed both at wrist $O_1$ and at tip $A_0$ of the hand (Fig. 7). Furthermore, the hip-torque limit of the lower leg is exceeded between knots. In brief, the minimum energy cost generates fast varying torques which are incompatible in the present case with one-sided contact conditions.

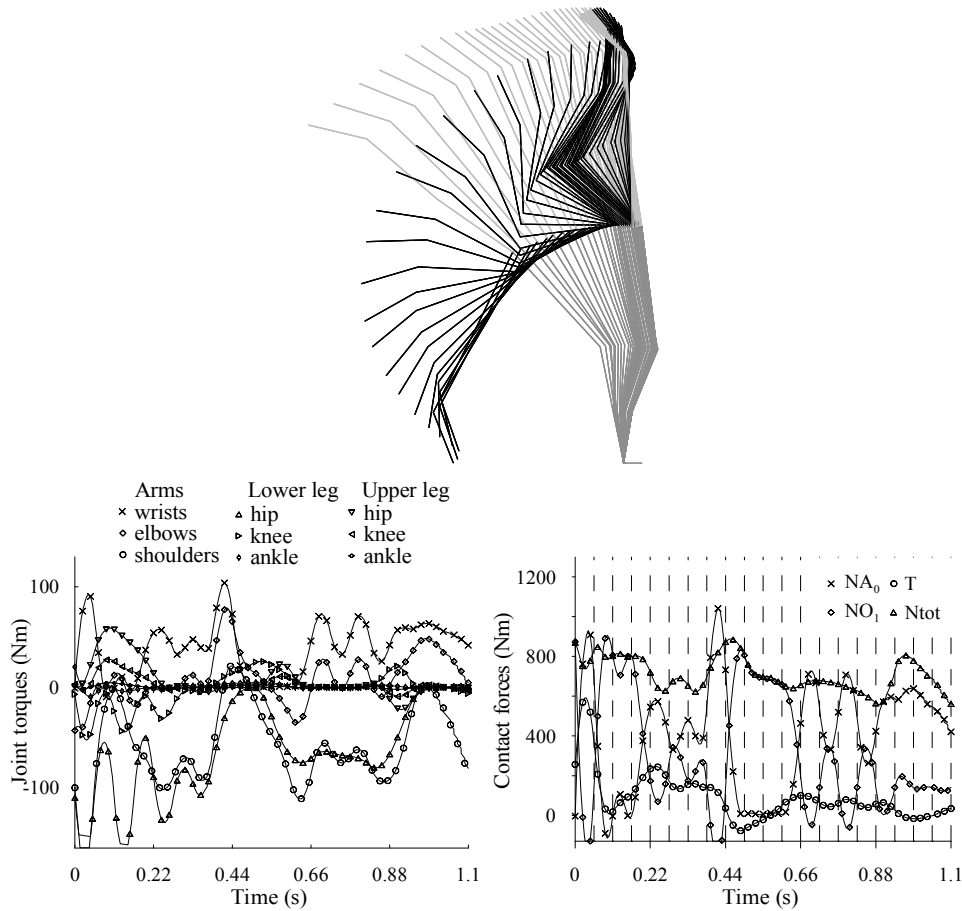

Fig. 7. *Stick diagram of movement, and time charts of driving torques and contact forces generated by a purely energetic criterion* $(\alpha = 0)$.

Through the above examples, we can conclude that the energetic cost creates movements with little internal gesticulation and fast varying actuating torques. Quite the reverse, the sthenic cost generates more gesticulation while driving torques show small variations.

In two further simulations we introduce state constraints in order for the optimized motions to mimic the gymnast movement. Since the gymnast performs slight leg flexions, we limit the knee flexions to 30 degrees as given in (33). Moreover, the initial hand-foot distance is no longer free but is fixed to 0.95m, the value observed in the real movement (Fig. 2).

Figure 8 shows that the kinematic differences between both optimized motions are attenuated by the new constraints. However, there is still more leg flexion generated by the sthenic criterion than its nearly energetic counterpart at beginning and end of movements.

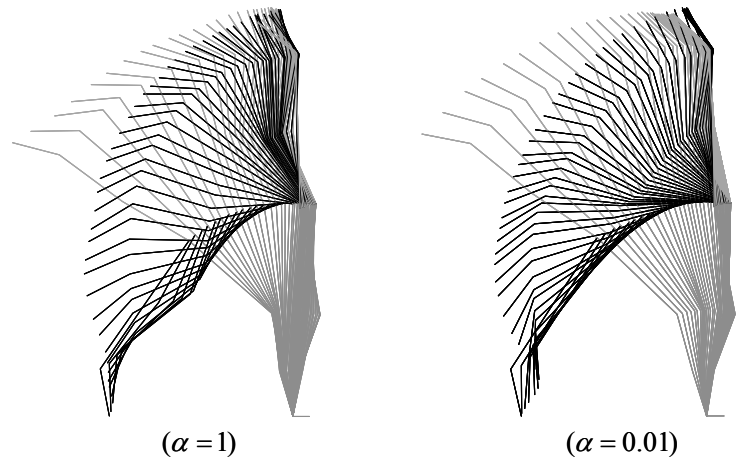

($\alpha = 1$)    ($\alpha = 0.01$)

Fig. 8. *Stick diagram of constrained motions mimicking the gymnast movement.*

In the same way, the main differences shown previously in Fig. 6 reappear in Fig. 9: there are more variations for joint torques and contact forces in the energetic case than in the sthenic one. Furthermore, in the minimum energy case, the lower leg hip-torque limit is not respected between knots and the unilaterality condition is slightly infringed at the beginning of the motion.

To finish, we make a comparison between the dynamics of both the simulated movement and the movement performed by the gymnast. The torques generated by the gymnast were computed using inverse dynamic analysis. The kinematics was determined by means of a video recording system, data filtering and numerical differentiations techniques to compute joint velocities and accelerations.

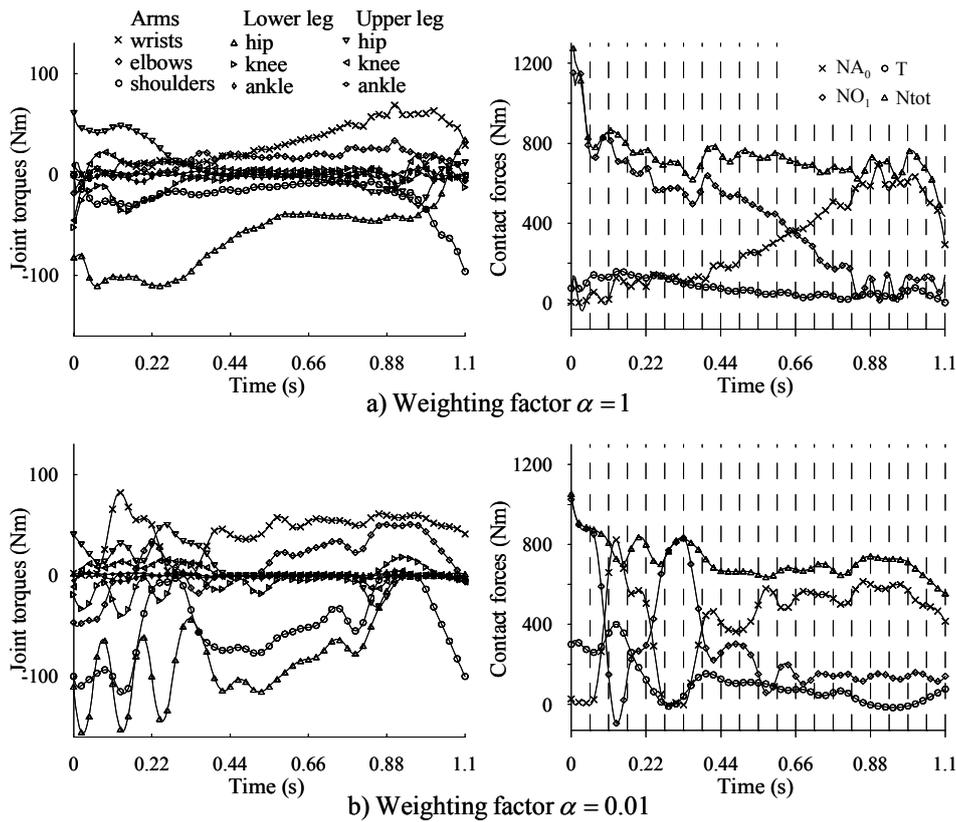

Fig. 9. *Time charts of driving torques and contact forces generated by the sthenic ($\alpha = 1$) and the nearly energetic criterion ($\alpha = 0.01$) with limited flexion range at knees, and with an initial foot-hand distance equal to that of the real movement performed by the gymnast.*

Fig. 10 shows that torques at hips and knees exhibit moderate extremal values and few variations. Arm driving torques are more fluctuating than the previous ones and reach sizeable values showing that the gymnast uses his arms very much to control his movement. The comparison between these results with their counterparts shown in Fig. 9 is not obvious. Nevertheless, in the nearly energetic case ($\alpha = 0.01$), one can see that torques at wrists and shoulders take important fast-varying values, especially at the beginning, which is fairly similar to the situation observed in Fig. 10. More striking similarities appear in the global results shown in Table 3: the actuating work and the initial kinetic energy are roughly of the same order in the simulated energetic case and in the real movement. But, in this respect, one can wonder why the mechanical energy expended in the latter case is less than in the former which represents an idealized situation. This is probably due to the fact that real actuating torques are undervalued through the numerical treatment undergone by the kinematic data. Indeed, data filtering and numerical differentiations yield velocities and even more accelerations of which peak values are attenuated. This certainly affects the computed dynamics which is found weaker than it is really.

Although the above comparisons are not quite clearly definite, they show that the dynamics of the physical movement has real similarities with the minimum-energy optimal motion. These results show also the need for improving both the accuracy of experimental and numerical techniques required to carry out dynamic analysis of human movements, and the precision of optimization methods which provide only local suboptimal solutions.

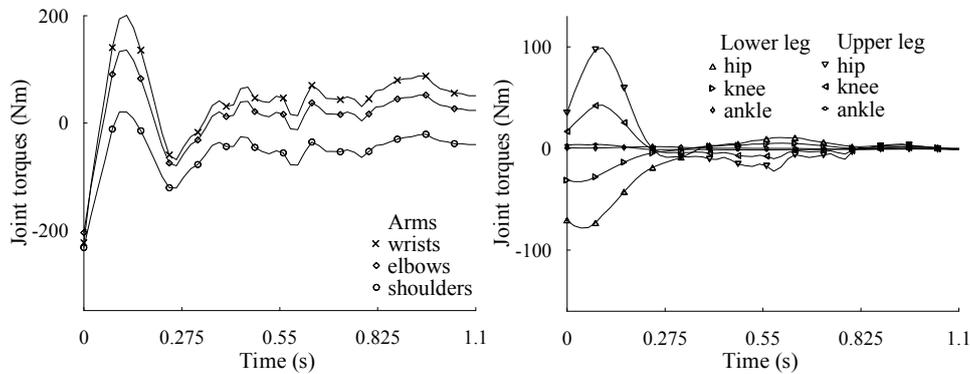

Fig. 10. *Time charts of actuating torques exerted by the gymnast during the raising phase of the floor handstand.*

| $\alpha$ | Minimal cost $J$ (adimensional) | Actuating effort $J_s$ (adimensional) | Actuating work $J_e$ (J) | Initial kinetic energy (J) |
|---|---|---|---|---|
| 1 | 12.8 | 12.8 | 375 | 197 |
| 0.01 | 1.21 | 23.9 | 172 | 175 |
| Movement performed by the gymnast | | | 144 | 157 |

Table 3. *Minimal costs obtained with two additional constraints: foot-hand distance is fixed at 0.95m, and knee flexion is limited to 30°.*

## 6. Concluding remarks

Optimal dynamic synthesis is a significant means to gain an insight into the way kinematic organization and dynamic coordination of human movements are linked together. Quite different results may be found according to the dynamic cost minimized. Indeed, the analysis of numerical simulations showed that the choice of a performance criterion is an ambiguous question because human movements are in the heart of a compromise between the least energy expended and the least effort put in.

We paid a special attention to dealing with the energy cost. Especially, the effect of such a criterion was emphasized, and the problem of its implementation was explicitly clarified. As the minimum energy cost generates bang-off-bang actuating torques, we underline that it requires an accurate parametric optimization method to reveal the fast and complex variations of suboptimal inputs which result. Moreover, it appeared that a purely energy cost could be not satisfactory for generating movements with one-sided contact conditions which might be infringed as during the raising phase of floor handstands.

A better approach is based on considering a mixed criterion associating actuating effort with driving work. This way of tackling dynamic movement synthesis offers a helpful adaptability of the optimization process. In numerical simulations, the progressive change from one cost to the other reveals that the movement is controlled in quite different ways, which results in noticeable modifications of its kinematics. Moreover, while the criterion with predominant energy generates movements with kinematic aspects similar to the real handstand performed by a gymnast, the movement produced by the sthenic criterion exhibits quite dissimilar features. Nevertheless, this movement characterized by smooth actuating torques proved to be easier to execute by novice gymnasts. Conversely, movements governed by a minimum energy cost are harder to perform probably for the reason that they need to control accurately successive brief impulses of actuating torques.

## Appendix

- *Solving an elementary maximization problem*

The presentation which follows generalizes to the cases $a \neq -b$ and $\beta \geq 0$ (see below) a result originally established in a somewhat different form in [33] and in [23].

In (17), setting $a = \tau_i^{\min}$ and $b = \tau_i^{\max}$, and removing the subscript *i*, we want to answer the question:

– Find $u \in [a,b]$, $a < 0$ and $b > 0$, solution of the maximization problem:

$$\underset{\alpha \in \mathfrak{R}, \beta \geq 0}{\text{Max}} (\alpha u - \beta |u|).$$

First, assuming that $\beta \neq 0$, we define the function

$$\varphi(u) = \begin{cases} a_1 u, \ a_1 = \alpha + \beta, & \text{if } u \leq 0 \\ a_2 u, \ a_2 = \alpha - \beta, & \text{if } u > 0 \end{cases}. \tag{A1}$$

Next, we consider the successive cases

- Case 1: $\alpha < -\beta$

  Then $a_2 < a_1 < 0$, and the function $\varphi$ reaches its maximum on $[a,b]$ at $a$ (see Fig. A1). Thus the solution is $u = a$.

  If $\alpha/\beta = -1$, then $a_1 = 0$ and $u \in [a,0]$.

- Case 2: $-\beta < \alpha < \beta$

  Then $a_2 < 0$, $a_1 > 0$, and the maximum of $\varphi$ is reached for $u = 0$.

- Case 3: $\beta < \alpha$

  Then $a_1 > a_2 > 0$, and $\varphi$ reaches its maximum for $u = a$.

  If $\alpha/\beta = 1$, then $a_2 = 0$ and $u \in [0,b]$.

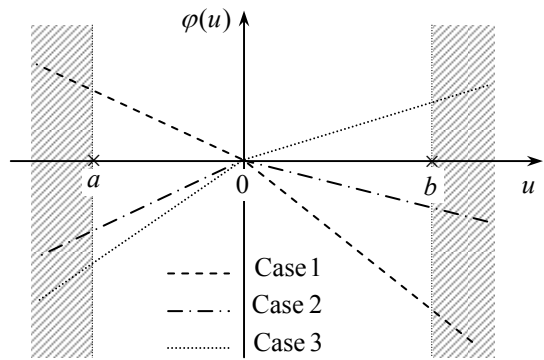

Fig. A1. *Variations of the function $\varphi$ over the interval $[a,b]$*

These results can be summarized using the *dead zone function* (see [23]) generalized here as the following asymmetrical function of the three variables $\alpha/\beta$, $a$ and $b$ (Fig. A2):

$$u = \text{dez}(\alpha/\beta, a, b) \begin{cases} = a & \text{if } \alpha/\beta < -1 \\ \in [a,0] & \text{if } \alpha/\beta = -1 \\ = 0 & \text{if } -1 < \alpha/\beta < 1 \\ \in [0,b] & \text{if } \alpha/\beta = 1 \\ = b & \text{if } 1 < \alpha/\beta \end{cases} \quad (A2)$$

If $\beta = 0$, then the maximizing element $u$ of $\varphi$ in (A1) appears as the sign function:

$$u = a\,\text{sign}(\alpha). \quad (A3)$$

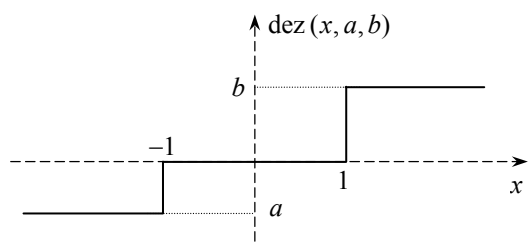

Fig. A2. *Graph of an asymmetrical dead zone function* dez(*x,a,b*)

- *Biometric data*

Reference [34] was used to compute the biometric data of the expert gymnast.

| Body segment | Mass (kg) | Length (m) | Local abscissa of CoG $C_i$ (m) | Moment of inertia at $C_i$ (kg.m$^2$) |
|---|---|---|---|---|
| Hand | 0.408 | 0.093 | 0.046 | 0.005 |
| Fore-arm | 3.808 | 0.201 | 0.114 | 0.016 |
| Arm | 2.176 | 0.263 | 0.150 | 0.014 |
| Trunk-head | 39.30 | 0.568 | 0.193 | 3.213 |
| Thigh | 6.800 | 0.434 | 0.188 | 0.134 |
| Shin | 3.162 | 0.372 | 0.161 | 0.040 |
| Foot | 0.986 | 0.126 | 0.063 | 0.035 |

Table A1. *Gymnast biometric data.*